\documentclass{iopart}
\usepackage{iopams}
\usepackage{euscript,amssymb,epsfig}

\usepackage{amsfonts,latexsym}
\usepackage{euscript,amssymb}
\usepackage{epsfig}

\usepackage{graphicx}

\newcommand{\be}[1]{\begin{equation}\label{#1}}
\newcommand{\ee}{\end{equation}}
\newcommand{\ba}[1]{\begin{eqnarray}\label{#1}}
\newcommand{\ea}{\end{eqnarray}}
\newcommand{\rf}[1]{(\ref{#1})}
\newcommand{\nn}{\nonumber}

\begin{document}

\title{Non-relativistic limit of Randall-Sundrum model: solutions, applications and constraints}

\author{Maxim Eingorn, Alexandra Kudinova and Alexander Zhuk}

\address{Astronomical Observatory and Department of
Theoretical Physics, Odessa National University, Street Dvoryanskaya 2, Odessa 65082, Ukraine}

\eads{\mailto{maxim.eingorn@gmail.com}, \mailto{autumnforever1@gmail.com} and \mailto{ai.zhuk2@gmail.com}}


\begin{abstract}
In the Randall-Sundrum model with one brane, we found the approximate and exact solutions for gravitational potentials and accelerations of test bodies in these
potentials for different geometrical configurations. We applied these formulas for calculation of the gravitational interaction between two spheres and found the
approximate and exact expressions for the relative force corrections to the Newton's gravitational force. We demonstrated that the difference between relative force
corrections for the approximate and exact cases increases with the parameter $l$ (for the fixed distance $r$ between centers of the spheres). On the other hand, this
difference increases with decreasing of the distance between the centers of the spheres (for the fixed curvature scale parameter $l$). We got the upper limit for the
curvature scale parameter $l\lesssim 10\, \mu$m. For these values of $l$, the difference between the approximate and exact solutions is negligible.
\end{abstract}


\maketitle


\section{\label{sec:1}Introduction}

\setcounter{equation}{0}

The idea of the multidimensionality of our Universe  has been attracting continuous interest for many years. It takes its origin from the pioneering papers by Th. Kaluza
and O. Klein \cite{KK}, and now the most self-consistent modern theories of unification such as superstrings, supergravity and M-theory are constructed in spacetime with
extra dimensions \cite{Polchinski}. In Kaluza-Klein models, our spacetime is effectively four-dimensional due to compactness and smallness of the extra dimensions
(internal spaces). The size of the extra dimensions is restricted by the electroweak scales $10^{-17}$ cm. However, our spacetime can be effectively four-dimensional
even in the case of infinite extra dimensions. This interesting scenario is realized in recently proposed brane world models (see, e.g., the reviews \cite{Rub,Barv}).
Here, matter fields from the Standard Model are  trapped to a three-dimensional submanifold (brane) embedded in the fundamental multidimensional space (bulk), but
gravity may move in the bulk. Localization of massless gravitons on a brane results in effective four-dimensional Einstein gravity in the low energy limit. Certainly,
large and infinite extra dimensions are potentially detectable. This was one of the main reasons for the great interest in this scenario. Therefore, it is very important
to suggest experiments which can reveal such extra dimensions.

In our paper, we consider the scenario that was first proposed in \cite{RS2}. Here, the brane is embedded in the five-dimensional anti-DeSitter spacetime, which allows
the extra dimension to be infinite. A negative bulk cosmological constant $\Lambda_5$ and a brane tension $\sigma$ are fine tuned to each other. Clearly, this is a very
simplified scenario. However, it gives a possibility to reveal some general features of the brane world models, in particular, the localization of the massless graviton
on the brane that restores the Newtonian limit on the brane at large distances from the gravitating matter source. It was shown \cite{RS2} that at distances greater than
a curvature scale of anti-DeSitter spacetime $r\gg l\sim |\Lambda_5|^{-1/2}$, the gravitational potential takes an approximate form with a cubic additive $\sim 1/r^3$ to
the usual Newtonian potential $\sim 1/r$. This approximate solution is much simpler than the exact one that makes the investigation of the effects of the extra dimension
much easier. In some papers (see, e.g., \cite{approx1,approx2}) this approximation was used to calculate the gravitational interaction between gravitating test bodies of
different geometrical form.
But we should analyze the difference between the approximate and exact solutions to find out where the application of the approximate solution is appropriate. This is
one of the main motivations of this work. To perform such analysis, we obtain two types of solutions (approximate and exact) for gravitational potentials and
accelerations of test bodies in these potentials for different geometrical configurations. Then, we apply these formulas to the most interesting for experiments case of
gravitational interaction between two massive spheres. We calculate approximate and exact corrections to the Newton's gravitational force and show that the difference
between relative force corrections for the approximate and exact cases increases with the parameter $l$ (for the fixed distance $r$ between centers of the spheres). On
the other hand, this difference increases with decreasing of the distance between the centers of the spheres (for the fixed curvature scale parameter $l$). The relative
force corrections also allow us to get the experimental constraint on the curvature scale parameter: $l\lesssim 10\, \mu$m. To get it, we use the results of the
table-top inverse square law experiments for the measurements of the Newton's gravitational constant. This is one of the main results of our paper.

The paper is structured as follows. In section 2 we describe briefly the Randall-Sundrum model with one brane. Here, we consider non-relativistic limit of this model and
present approximate and exact solutions for the gravitational potential on the brane. These formulas are applied to some practical problems in section 3 to get
approximate and exact expressions for the gravitational potential and acceleration of a point mass for these problems. In section 4 we investigate the gravitational
interaction of two spherical shells. Then, in section 5 we compare the relative corrections to the gravitational force between two spheres in approximate and exact
cases. Here, we also get the constraint on the curvature scale parameter in the Randall-Sundrum model. A brief discussions of the obtained results is presented in the
concluding section 6.


\section{\label{sec:2}Non-relativistic limit of Randall-Sundrum model}

\setcounter{equation}{0}

The one-brane Randall-Sundrum metrics is \cite{RS2}
\be{2.1}
ds^2=\exp\left(-\frac{2|\xi|}{l}\right)\eta_{\mu\nu}dx^{\mu}dx^{\nu}-d\xi^2\ ,
\ee
where $\eta_{\mu\nu}$ is the flat four-dimensional spacetime metrics and the parameter $l$ is defined via the 5-dimensional cosmological constant:
\be{2.2}
\Lambda_5=-\frac6{l^2}\, ,
\ee
i.e. $l$ is the curvature scale of 5-dimensional anti-DeSitter spacetime. The brane is embedded in this spacetime at $\xi=0$ and has fine tuned tension
\be{2.3} \sigma=\frac{3c^4}{4\pi G_5 l}\, , \ee
where $G_5$ is the 5-dimensional gravitational constant. In one-brane Randall-Sundrum scenario, the extra dimensions is infinite: $\xi \in (-\infty,+\infty)$.

Now, we want to probe this model with the help of the gravitational terrestrial experiments, e.g., the inverse square law  experiments. Certainly, this is the case of
non-relativistic limit of the model. In this limit, we need to get the gravitational potential $\varphi (r)$ on the brane.
Following, e.g., the calculations in \cite{Barv}, we obtain
\be{2.4}
\varphi(r)=-\frac{G_5 m}{r l}-\frac{G_5 m}{r}\int\limits_0^{\infty}d\tilde m \varphi^2_{\tilde m}(l)\exp(-\tilde m r)\, ,
\ee
where $r$ is the magnitude of the radius vector on the brane and
\be{2.5}\varphi_{\tilde m}(l)=\left(\frac{\tilde m l}{2}\right)^{1/2}\frac{Y_1(\tilde m l)J_2(\tilde m l)-J_1(\tilde m l)Y_2(\tilde m l)}{\left(J^2_1(\tilde m
l)+Y^2_1(\tilde m l)\right)^{1/2}}\, , \ee
where $J$ and $Y$ are Bessel functions of the first and second kinds, respectively. In the short and long distance limits the equation \rf{2.4} reads, respectively:
\be{2.6} \varphi(r)\approx-\frac{G_5 m}{rl}-\frac{G_5 m}{r}\frac{1}{\pi r}
\approx-\frac{G_5 m }{\pi r^2}\, ,\  \ \ \ r\ll l\,
\ee
and
\be{2.7}
\varphi(r)\approx-\frac{G_N m}{r}\left(1+\frac{\alpha}{r^2}\right)\ , \ \ \ r\gg l\, ,
\ee
where we have introduced the Newton's gravitational constant
\be{2.8}
G_N=\frac{G_5}{l}
\ee
and the parameter\footnote{\label{alpha}It is worth noting that in the pioneering paper \cite{RS2} $\alpha = l^2$. The brane-bending effect \cite{GT} gives
$\alpha=2l^2/3$. In the paper \cite{JKP}, the authors have pointed out that different schemes of regularization result in different values of $\alpha$. In our paper we
follow calculations in \cite{Barv}, where $\alpha=l^2/2$.}
\be{2.9}
\alpha = \frac{l^2}{2}\, .
\ee
Obviously, \rf{2.6} corresponds to the strong deviation from the Newtonian gravity but the formula \rf{2.7} describes the smooth transition to the Newtonian limit. The
exact expression \rf{2.4} (the solid line) and its asymptotes \rf{2.6} and \rf{2.7} (the short-dashed and long-dashed lines, respectively) are depicted on figure 1.
Here, we introduce the dimensionless distance argument $\eta=r/l$ and dimensionless potentials $\tilde \varphi (\eta) =\varphi (r)/(G_Nm/l)$.

\begin{figure}
  \center
    \includegraphics[width=3.0in,height=2.2in]{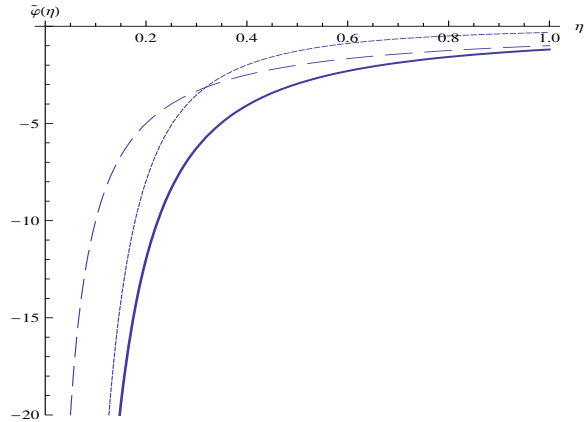}\\
\caption {The gravitational field potential \rf{2.4} (the solid line) and its asymptotes \rf{2.6} and \rf{2.7} (the short-dashed and long-dashed lines, respectively).}
\end{figure}


\section{\label{sec:3}Applications}

\setcounter{equation}{0}

Now, we want to apply the obtained formulas to terrestrial gravitational experiments. Obviously, the gravitational field on the Earth should not considerably differ from
the Newtonian one. Therefore, we should use either the exact expression \rf{2.4} or the approximate formula \rf{2.7}. Therefore, we shall get two classes of solutions:
exact and approximate, respectively. Obviously, the approximate formula \rf{2.7} looks much more simple. However, it is necessary to check the deviation of expressions
based on it from the exact ones for real gravitational experiments. This is one of the main aims of the paper.

\subsection{Infinitesimally thin shell}

Let us consider first an infinitesimally thin shell of the mass $m=4\pi R^2\sigma$, where $R$ and $\sigma$ are the radius and the surface mass density of the shell.
Then, the gravitational potential of this shell in a point with the radius vector $\bf r$ (from the center of the shell) for the approximate solution is
\be{3.1} \varphi(r>R)=-\frac{G_N m}{r}\left[1+\frac{\alpha}{r^2-R^2}\right]\ee
and
\be{3.2}
\varphi(r<R)=-\frac{G_N m}{R}\left[1+\frac{\alpha}{R^2-r^2}\right]\, .
\ee
Obviously, these expressions are divergent when $r\rightarrow R$: $\varphi(r\gtrless R)\rightarrow-G_N m\alpha/\left[2R^2|r-R|\right] \rightarrow -\infty$. In the case
of the exact solution we get
\ba{3.3}
\fl \varphi (r>R)=-\frac{G_N m}{r}\left[1+\frac{l^2}{2R}\int\limits_0^{\infty}d\tilde m \left[\frac{Y_1(\tilde ml)J_2(\tilde ml)-J_1(\tilde ml)Y_2(\tilde
ml)}{(J_1^2(\tilde ml)+Y_1^2(\tilde ml))^{1/2}}\right]^2\sinh(\tilde m R)e^{-\tilde m r}\right]\nn \\
\phantom{}
\ea
and
\ba{3.4}
\fl \varphi(r<R)=-\frac{G_N m}{R}\left[1+\frac{l^2}{2r}\int\limits_0^{\infty}d\tilde m \left[\frac{Y_1(\tilde ml)J_2(\tilde ml)-J_1(\tilde ml)Y_2(\tilde
ml)}{(J_1^2(\tilde ml)+Y_1^2(\tilde ml))^{1/2}}\right]^2\sinh(\tilde m r) e^{-\tilde m R} \right].\nn \\
\phantom{}
\ea
It is not difficult to verify that these exact solutions are also divergent when $r\to R$.

Formulas \rf{3.2} and \rf{3.4} demonstrate that inside of the shell the gravitational potential is not a constant. Thus, a test body undergoes an acceleration (see
\rf{3.6} and \rf{3.8} below) in contract to the Newtonian case, i.e. the Birkhoff's theorem is violated. The acceleration outside and inside of the shell is
\be{3.5}
-\frac{d\varphi}{dr}(r>R)=-\frac{G_N m}{r^2}\left[1+\alpha\frac{3r^2 - R^2}{(r^2-R^2)^2}\right]
\ee
and
\be{3.6}
-\frac{d\varphi}{dr}(r<R)=\frac{G_N m}{R}\frac{2\alpha r}{(R^2-r^2)^2}\, ,
\ee
which is divergent when $r\rightarrow R$: $-\frac{d\varphi}{dr}(r\gtrless R)\rightarrow \mp \frac{G_N m\alpha}{2R^2(r-R)^2} \rightarrow \mp \infty$ (the upper and lower
signs correspond to $r>R$ and $r<R$, respectively).

In the case of exact solutions we have
\ba{3.7}
\fl &-&\frac{d\varphi}{dr} (r>R)=-\frac{G_N m}{r^2}\\
\fl &-&\frac{G_N m l^2}{2Rr^2}\int\limits_0^{\infty}d\tilde m \left[\frac{Y_1(\tilde ml)J_2(\tilde ml)-J_1(\tilde
ml)Y_2(\tilde ml)}{(J_1^2(\tilde ml)+Y_1^2(\tilde ml))^{1/2}}\right]^2\sinh(\tilde m R)(1+\tilde m r)e^{-\tilde m r} \nonumber
\ea
and
\ba{3.8}
\fl &-&\frac{d\varphi}{dr}(r<R)\\
\fl &=&\frac{G_N ml^2}{2Rr^2}\int\limits_0^{\infty}d\tilde m \left[\frac{Y_1(\tilde ml)J_2(\tilde ml)-J_1(\tilde ml)Y_2(\tilde
ml)}{(J_1^2(\tilde ml)+Y_1^2(\tilde ml))^{1/2}}\right]^2(\tilde m r\cosh(\tilde m r)-\sinh(\tilde m r))e^{-\tilde m R}. \nonumber \ea
The exact solutions \rf{3.7} and \rf{3.8} are also divergent for $r\rightarrow R$.

\subsection{Spherical shell of finite thickness}

Here, we consider a spherical shell of the inner radius $R_1$ and the outer radius $R_2$ and the mass $m=(4\pi \rho/3)\left(R_2^3-R_1^3\right)$ with a constant volume
density $\rho$\footnote{\label{limit}It is clear that the limit $R_2 \rightarrow R_1$ is incorrect because, for constant $\rho$, it results in vanishing $m$  and, vice
versa, for fixed $m$ the volume density $\rho$ goes to infinity. Therefore, such a naive limit does not provide us the correct transition to the formulas from the
previous subsection.}. For this geometry, the approximate gravitational potential reads
\be{3.9}
\fl \varphi(r>R_2)=-\frac{G_N m}{r}\left[1+\frac{3\alpha}{R_2^3-R_1^3}\left(R_1-R_2+\frac{r}{2}\ln\frac{(r+R_2)(r-R_1)}{(r-R_2)(r+R_1)}\right)\right]
\ee
and
\be{3.10}
\fl \varphi(r<R_1)=-2\pi G_N\rho\left(R_2^2-R_1^2+\alpha\ln\frac{R_2^2-r^2}{R_1^2-r^2}\right)\, .
\ee
These expressions are logarithmically divergent in the vicinity of $R_1$ and $R_2$: $\varphi(r>R_2)\rightarrow \frac{3G_N m\alpha}{2\left(R_2^3-R_1^3\right)}\ln(r-R_2)
\rightarrow -\infty$ for $r\rightarrow R_2$ and $\varphi(r<R_1)\rightarrow 2\pi G_N\rho\alpha\ln(R_1-r) \rightarrow -\infty $ for $r\rightarrow R_1$.

For the exact solution we get
\ba{3.11}
\fl &{}& \varphi(r>R_2)=-\frac{G_N m}{r}\left\{1+\frac{l^2}{2}\int\limits_0^{\infty}d\tilde m \left[\frac{Y_1(\tilde ml)J_2(\tilde ml)-J_1(\tilde ml)Y_2(\tilde
ml)}{(J_1^2(\tilde ml)+Y_1^2(\tilde ml))^{1/2}}\right]^2\frac1{\tilde m^2}e^{-\tilde m r}\right.\nonumber\\
\fl &{}&\left.\times\frac3{(R_2^3-R_1^3)}\left[R\tilde m\cosh(\tilde m R)-\sinh(\tilde mR)\right]|_{R_1}^{R_2}\right\}
\ea
and
\ba{3.12}
\fl &{}&\varphi(r<R_1)=-4\pi G_N\rho\left\{\frac{R^2}{2}-\frac{l^2}{2r}\int\limits_0^{\infty}d\tilde m \left[\frac{Y_1(\tilde ml)J_2(\tilde ml)-J_1(\tilde
ml)Y_2(\tilde ml)}{(J_1^2(\tilde ml)+Y_1^2(\tilde ml))^{1/2}}\right]^2\frac1{\tilde m^2}\sinh(\tilde m r)\right.\nonumber\\
\fl &{}&\times\left.(\tilde m R +1)e^{-\tilde m R}\phantom{\int}\right\}|_{R_1}^{R_2}\, .
\ea
In contrast to the approximate formulas \rf{3.9} and \rf{3.10}, these exact expressions are convergent in the limits $r \rightarrow R_1,R_2$.

The acceleration of a test body outside and inside of the shell is
\be{3.13}
\fl -\frac{d\varphi}{dr}(r>R_2)=-\frac{G_N
m}{r^2}\left[1+\frac{3\alpha}{R_2^3-R_1^3}\left(R_1-R_2+r^2\frac{(R_2-R_1)(r^2+R_1R_2)}{(r^2-R_1^2)(r^2-R_2^2)}\right)\right]\quad
\ee
and
\be{3.14}
\fl -\frac{d\varphi}{dr}(r<R_1)=4\pi G_N\rho\alpha r\frac{R_2^2-R_1^2}{(R_2^2-r^2)(R_1^2-r^2)}\, .
\ee
These formulas are divergent in the limits $r \rightarrow R_1,R_2$: $-\frac{d\varphi}{dr}(r>R_2)\rightarrow -\frac{3G_N m\alpha}{2\left(R_2^3-R_1^3\right)(r-R_2)}
\rightarrow -\infty$ for $r\rightarrow R_2$ and $-\frac{d\varphi}{dr}(r<R_1)\rightarrow \frac{2\pi G_N\rho\alpha}{R_1-r} \rightarrow +\infty$ for $r\rightarrow R_1$.

For exact solutions we obtain
\ba{3.15}
\fl &-&\frac{d\varphi}{dr}(r>R_2)=-\frac{G_N m}{r^2}-\frac{G_N ml^2}{2r^2}\int\limits_0^{\infty}d\tilde m \left[\frac{Y_1(\tilde ml)J_2(\tilde ml)-J_1(\tilde
ml)Y_2(\tilde
ml)}{(J_1^2(\tilde ml)+Y_1^2(\tilde ml))^{1/2}}\right]^2\nonumber\\
\fl &\times&\frac1{\tilde m^2}e^{-\tilde m r}\left(1+\tilde m r\right)\frac3{(R_2^3-R_1^3)}\left[R\tilde m\cosh(\tilde m R)-\sinh(\tilde mR)\right]|_{R_1}^{R_2}
\ea
and
\ba{3.16}
\fl &-&\frac{d\varphi}{dr}(r<R_1)=-\frac{2\pi G_N\rho l^2}{r^2}\int\limits_0^{\infty}d\tilde m \left[\frac{Y_1(\tilde ml)J_2(\tilde ml)-J_1(\tilde ml)Y_2(\tilde
ml)}{(J_1^2(\tilde
ml)+Y_1^2(\tilde ml))^{1/2}}\right]^2\nonumber\\
\fl &\times& \frac1{\tilde m^2}(\tilde m r\cosh(\tilde m r)-\sinh(\tilde m r)) (\tilde m R+1)e^{-\tilde m R}|_{R_1}^{R_2}\, .
\ea
These integrals are divergent in the vicinity of $R_1$ and $R_2$.

\subsection{Sphere}

Obviously, all formulas for a sphere of the radius $R$ and the mass $m=4\pi \rho R^3 /3$ with a constant volume density $\rho$ can be easily obtained from the equations
\rf{3.9}, \rf{3.11}, \rf{3.13} and \rf{3.15}  with the help of the evident substitutions: $R_1=0$ and $R_2\equiv R$.

\section{\label{sec:4}Gravitational interaction of two spherical shells}

\setcounter{equation}{0}

Let us consider now two spherical shells with radii $R_2>R_1$ and the mass $m=(4\pi \rho/3)\left(R_2^3-R_1^3\right)$ for the first shell and radii $R'_2>R'_1$ and the
mass $m'=(4\pi \rho'/3)\left(R_2^{'3}-R_1^{'3}\right)$ for the second shell. Then, the potential energy of gravitational interaction between these shells for the
approximate solution reads
\ba{4.1}
\fl U(r)&=&-\frac{G_N mm'}{r}-\frac{2\pi^2G_N\rho\rho'\alpha}{r}
\left\{\left[-\frac1{12}r^4+\frac1{2}r^2\left(R'^2+R^2\right)+\frac1{4}\left(R'^2-R^2\right)^2\right]\right.\nn \\
\fl &\times& \ln\frac{r^2-(R'+R)^2}{r^2-(R'-R)^2}
+\frac{2}{3}r\left[R'^3\ln\frac{(r+R)^2-R'^2}{(r-R)^2-R'^2}+R^3\ln\frac{(r+R')^2-R^2}{(r-R')^2-R^2}\right]\nn \\
\fl &-&\left.\frac1{3}r^2R'R-R'^3R-R'R^3\right\}|_{R=R_1}^{R=R_2}|_{R'=R'_1}^{R'=R'_2}\,  ,
\ea
where $r\geqslant R_2+R'_2$ is the distance between the centers of the shells and $f(R,R')|_{R=R_1}^{R=R_2}|_{R'=R'_1}^{R'=R'_2}=f(R_2,R'_2) -f(R_2,R'_1) -f(R_1,R'_2)
+f(R_1,R'_1)$.

In the case of the exact solution we get
\ba{4.2}
\fl U(r)&=&-\frac{G_N m m'}{r}-\frac{G_N m m'}{r}\, \frac9{(R_2^3-R_1^3)(R_2^{'3}-R_1^{'3})} \nonumber\\
\fl &\times& \frac{l^2}{2}\int\limits_0^{\infty}d\tilde m \left[\frac{Y_1(\tilde ml)J_2(\tilde ml)-J_1(\tilde ml)Y_2(\tilde ml)}{(J_1^2(\tilde ml)+Y_1^2(\tilde
ml))^{1/2}}\right]^2 \frac1{\tilde m^5}e^{-\tilde m r}\nonumber\\
\fl &\times& \left[\tilde m R\cosh(\tilde m R)-\sinh(\tilde m R)\right]|_{R_1}^{R_2}\cdot\left[\tilde m R\cosh(\tilde m R)-\sinh(\tilde m R)\right]|_{R'_1}^{R'_2}\, .
\ea
The additional analysis shows that both of these expressions \rf{4.1} and \rf{4.2} are convergent in the limit $r\rightarrow R_2+R'_2$.

With the help of these formulas, we can obtain the absolute value of the gravitational force between two shells:
\be{4.3}
F(r)=\frac{dU}{dr} = \frac{G_N m m'}{r^2}(1+\delta_F)\, ,
\ee
where $\delta_F$ defines the relative deviation from the Newtonian expression $G_N m m'/r^2$. For the approximate and exact solutions we have, respectively:
\ba{4.4}
\fl \delta_F&=&-\frac{9\alpha}{8\left(R_2^3-R_1^3\right)\left(R_2'^3-R_1'^3\right)}\left\{\ln\frac{r^2-(R'+R)^2}{r^2-(R'-R)^2}
\left[-\frac1{4}r^4+\frac1{2}r^2\left(R'^2+R^2\right)\right.\right.\nn \\
\fl &-&\left.\left.\frac1{4}
\left(R'^2-R^2\right)^2\right]
-r^2R'R+R'^3R+R'R^3\right\}|_{R=R_1}^{R=R_2}|_{R'=R'_1}^{R'=R'_2}
\ea
and
\ba{4.5}
\fl \delta_F&=&\frac{9l^2}{2\left(R_2^3-R_1^3\right)\left(R_2^{'3}-R_1^{'3}\right)} \int\limits_0^{\infty}d\tilde m \left[\frac{Y_1(\tilde ml)J_2(\tilde
ml)-J_1(\tilde ml)Y_2(\tilde ml)}{(J_1^2(\tilde
ml)+Y_1^2(\tilde ml))^{1/2}}\right]^2\frac 1{\tilde m^5}e^{-\tilde m r}\nonumber\\
\fl &\times& (1+\tilde m r)\left[\tilde m R\cosh(\tilde m R)-\sinh(\tilde m R)\right]|_{R_1}^{R_2}
\cdot\left[\tilde m R'\cosh(\tilde m R')-\sinh(\tilde m R')\right]|_{R'_1}^{R'_2}\, .\nn \\
\fl
\ea
These relative corrections $\delta_F$ are also convergent in the limit $r\rightarrow R_2+R'_2$. In the limit of large separation between the shells $r\gg
R_{1,2},R'_{1,2}$ we obtain from \rf{4.4} $\delta_F=3\alpha/r^2 + O(1/r^3)$.

\section{\label{sec:5}Constraints}

\setcounter{equation}{0}

The obtained above formulas can be used for the experimental restrictions on the parameters of the model. In our case, it is the curvature scale $l$. To get it, we can
use the inverse square law experiments for two spheres. The potential energy of interaction and the gravitational force between two spheres follow from the previous
section with the help of the substitutions: $R_1=R'_1=0$ and $R_2\equiv R,R'_2=R'$. For example, the relative corrections to the gravitational force in approximate and
exact cases read, respectively:
\ba{5.1}
\fl \delta_F&=&-\frac{9\alpha}{8R^3R'^3}\left\{\ln\frac{r^2-(R'+R)^2}{r^2-(R'-R)^2} \left[-\frac1{4}r^4+\frac1{2}r^2\left(R'^2+R^2\right)-\frac1{4}
\left(R'^2-R^2\right)^2\right]\right.\nonumber\\
\fl&-& \left.r^2R'R+R'^3R+R'R^3\right\}
\ea
and
\ba{5.2}
\fl \delta_F&=&\frac{9l^2}{2R^3R^{'3}}\int\limits_0^{\infty}d\tilde m \left[\frac{Y_1(\tilde ml)J_2(\tilde ml)-J_1(\tilde ml)Y_2(\tilde ml)}{(J_1^2(\tilde
ml)+Y_1^2(\tilde ml))^{1/2}}\right]^2\frac 1{\tilde m^5}e^{-\tilde m r}(1+\tilde m r)\nonumber\\
\fl &\times& \left[\tilde m R\cosh(\tilde m R)-\sinh(\tilde m R)\right]\cdot\left[\tilde m R'\cosh(\tilde m R')-\sinh(\tilde m R')\right]\, .
\ea
We remind that $\alpha=l^2/2$ (see the equation \rf{2.9}). For definiteness, we shall use the parameters of the spheres from the Moscow Cavendish-type experiment
\cite{Moscow}: $R_1\approx 0.087$ cm for a platinum ball with the mass $m_1=59.25\times10^{-3}$ g, $R_2\approx0.206$~cm for a tungsten ball with the mass
$m_2=706\times10^{-3}$ g and the distance between their centers $r=0.3773$ cm.

It is clear that the use of the approximate solution for the gravitational interaction force makes the calculations much easier. But we should analyze the distinction
between the approximate and exact solutions to find out where the application of the approximate solution is appropriate. The difference between the approximate and
exact solutions for the relative force corrections is shown on figures 2 and 3. These figures demonstrate that the difference between these solutions increases with the
parameter $l$ (for the fixed distance $r$ between centers of the spheres) (figure 2). On the other hand, this difference increases with decreasing of the distance
between the centers of the spheres (for the fixed curvature scale parameter $l$) (figure 3).
\begin{figure}
  \center
    \includegraphics[width=3.0in,height=2.0in]{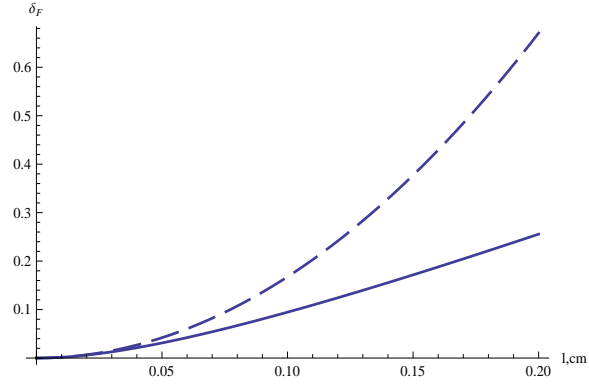}\\
\caption {Relative gravitational force corrections \rf{5.2} (the solid line) and \rf{5.1} (the dashed line) as functions of the curvature scale parameter $l$ in the case
of the distance between the centers of the spheres $r=0.3773$ cm.}
\end{figure}
\begin{figure}
  \center
    \includegraphics[width=3.0in,height=2.2in]{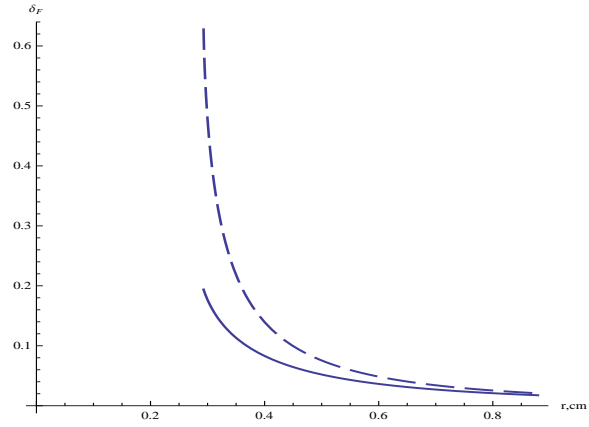}\\
\caption {Relative gravitational force corrections \rf{5.2} (the solid line) and \rf{5.1} (the dashed line) as functions of the distance between the centers of the
spheres in the case $l=10^{-1}$ cm.}
\end{figure}

Now, we want to estimate the curvature scale parameter $l$ with the help of our formulas \rf{5.1} and \rf{5.2}. To get it, we can use the value of the Newton's
gravitational constant $G_N$.
As it follows from figure 2 in the 'CODATA Recommended Values of the Fundamental Constants: 2006', the most precise values of $G_N$ were obtained in the University
Washington and the University Z\"urich experiments \cite{13,14}. They are $G_N / 10^{-11} {\mbox{m}}^3{\mbox{kg}}^{-1}{\mbox{s}}^{-2}=6.674\, 215\pm 0.000\, 092$ and
$6.674\, 252\pm 0.000\, 124$, respectively. The relative errors $\triangle G_N/G_N$ show the accuracy of the measurements of the gravitational constant in the inverse
square law experiments. If the correction $\delta_F$ due to the extra dimension is greater than these values, then we can detect the deviation from the Newton's law. Up
to now, there is no experimental evidence for such deviations. Therefore, the relation $|\triangle G_N/G_N | =\delta_F$ gives the upper limit for $\delta_F$. In turn,
the equations \rf{5.1} and \rf{5.2} show that $\delta_F \sim l^2$. Therefore, from these equations we can get the upper limit for $l$, substituting there for
definiteness values for the radii of the spheres and the separation between them from the Moscow experiment. Thus, for the Washington and Z\"urich experiments, in the
case of the approximate formula \rf{5.1} we get respectively 9.067$\, \mu$m and 10.527$\, \mu$m and in the case of the exact formula \rf{5.2} we obtain respectively
9.070$\, \mu$m and 10.531$\, \mu$m.

Of course, we get rather rough estimates for the upper limit of $l$. Anyway, we think that it gives more or less correct value of the order of magnitude of $l$ in the
Randall-Sundrum model with one brane: $l\lesssim 10\, \mu$m. Figure 2 shows that for such values of $l$ the difference between the approximate and exact formulas is
negligible. It is worth noting that close constraints were found in the table-top inverse square law experiments \cite{ISLex} and from astrophysical observations
\cite{BH1}-\cite{BH5}.

\section{Conclusion}

In our paper we have considered the one-brane Randall-Sundrum model. In the weak-field limit, we obtained the approximate and exact expressions for gravitational
potentials and accelerations of test bodies in these potentials for different geometrical configurations. Some of these approximate formulas were already known (see,
e.g., \cite{approx1,approx2}), but the exact ones were found for the first time. We applied these equations for calculation of the gravitational interaction between two
spherical shells of finite thickness that can be easily reduced to the case of spheres. Then, we found the approximate and exact expressions for the relative force
corrections to the Newton's gravitational force between two massive spheres. It is clear that the use of the approximate solution makes the calculations much easier. But
we should analyze the difference between the approximate and exact solutions to find out where the application of the approximate solution is appropriate. We found that
the difference between relative force corrections for the approximate and exact cases increases with the parameter $l$ (for the fixed distance $r$ between centers of the
spheres). On the other hand, this difference increases with decreasing of the distance between the centers of the spheres (for the fixed curvature scale parameter $l$).
Using the results of the table-top Cavendish-type experiments measuring the Newton's gravitational constant, from the equations for the relative force corrections we got
the upper limit for the curvature scale parameter $l\lesssim 10\, \mu$m in the Randall-Sundrum model. For these values of $l$, the difference between the approximate and
exact solutions is negligible.

\ack This work was supported in part by the "Cosmomicrophysics" programme of the Physics and Astronomy Division of the National Academy of Sciences of Ukraine.


\end{document}